\documentclass[11pt,letterpaper,copyedit]{article}
\usepackage{osajnl2,bm,amsmath,amssymb,graphicx}

\begin{document}
\newcommand{\ket}[1]{\left|#1\right\rangle}
\newcommand{\bra}[1]{\left\langle#1\right|}
\newcommand{\scl}[2]{\left\langle #1 | #2 \right\rangle}
\newcommand{\lm}{\lambda}

\title{Continuous-variable entanglement purification with atomic systems}
\author{Stojan Rebic$^{1}$, Stefano Mancini$^{2}$, Giovanna Morigi$^{3,4}$, and David Vitali$^{2}$}
\affiliation{$^{1}$Centre for Quantum Computing Technology, Macquarie University, Sydney, NSW 2109, Australia} \affiliation{$^{2}$Physics
Department, University of Camerino, I-62032 Camerino, Italy}
\affiliation{$^{3}$ Departament de F\'isica, Universitat Aut\`onoma de Barcelona, E-08193 Bellaterra, Spain}
\affiliation{$^4$ Theoretische Physik, Universit\"at des Saarlandes, D-66041 Saarbr\"ucken, Germany}

\begin{abstract}
We present a proposal for entanglement purification of the continuous-variable quantum state of two propagating optical fields. The scheme is based on each field interacting with a local node -atomic ensemble- whose internal collective excitation plays the role of an ancillary continuous variable resource. Entanglement purification is achieved by a dichotomic measurement, representing the required non-Gaussian element, which consists in detecting the presence or absence of collective excitations in the atomic ensemble. This scheme can be extended to networks, where the nodes are single trapped atoms, and constitutes an important building block for the implementation of a continuous-variable quantum repeater.
\end{abstract}

\date{\today}
\maketitle

\section{Introduction}

The property of entanglement lies at the heart of quantum information processing and quantum mechanics in general~\cite{Horodecki}. Continuous
variable (CV) entangled states are very attractive in this regard due to the relative easiness in their generation and manipulation, as compared
to their discrete counterpart~\cite{Braunstein05}. However, the degree of entanglement between two distant sites of a quantum network usually
decreases exponentially with the length of a connecting channel, calling for the implementation of some entanglement purification
procedures~\cite{Bennett96}. This involves purifying mixed entangled states, such as those that are obtained when the two halves of the
paradigmatic CV entangled state, the two-mode squeezed state \cite{Braunstein05}, have been distributed through noisy channels. For this
purpose, we recall that a procedure involving solely Gaussian local operations and classical communications does not lead to increase of the
initial entanglement~\cite{Eisert02}, and that non-Gaussian local operations must necessarily take part of any entanglement purification
protocol~\cite{Browne03}.

In this article we propose a scheme for increasing the CV entanglement of two traveling optical field: each beam is sent to a local node where it interacts in an appropriate way with atoms. In the limit in which the photonic field is coherent over a large number of atoms, a collective ensemble excitation is created. This excitation can be described by a continuous variable, a bosonic mode, and its interaction with the input optical mode realizes the first part of a purification protocol. The second important part is the non-Gaussian measurement, implemented through a dichotomic measure of presence or absence of atomic excitations in the ensemble. An alternative realization for the local node can be a single atom tightly confined by a harmonic potential, where the bosonic mode are the center-of-mass excitations, which are interfaced with light by means of the mechanical effects of atom-photon interactions~\cite{Zeng,Parkins,Morigi,Morigi2,Canizares}. In this case, the dichotomic measure consists in detecting whether or not the atom is in the ground state of the oscillator~\cite{Eschner}. We illustrate the efficiency of the protocol for the case when the optical modes are initially prepared in a two-mode thermal squeezed state, and show that the protocol is able to increase entanglement of the quantum state of the fields at the output with respect to the entanglement they possessed at the input.

We remark that, apart for quantum communication, the possibility to increase the available entanglement is helpful in various scenarios, such as for instance in quantum metrology. Here, entangled states allow to achieve the ultimate sensitivity allowed by the Heisenberg limit in phase-sensitive measurements, such as those employed in atomic clocks, and in weak force detection \cite{vittorio}.

The paper is organized as follows. In Section II, we introduce the theoretical model and discuss the basic assumptions. In Section III we use the characteristic function for the state of the system in order to evaluate the fidelity of the scheme. In Section IV we discuss the results, i.e., the amount of entanglement of the two optical fields at the output of the whole process, measured in terms of the teleportation fidelity, for all the possible conditional states. Concluding remarks and discussions are presented in Section V. The appendices report mathematical details, which complement the theory presented in Sec.~III and IV.


\section{The model}

\begin{figure}[t]
\begin{center}
\includegraphics[scale = 0.75]{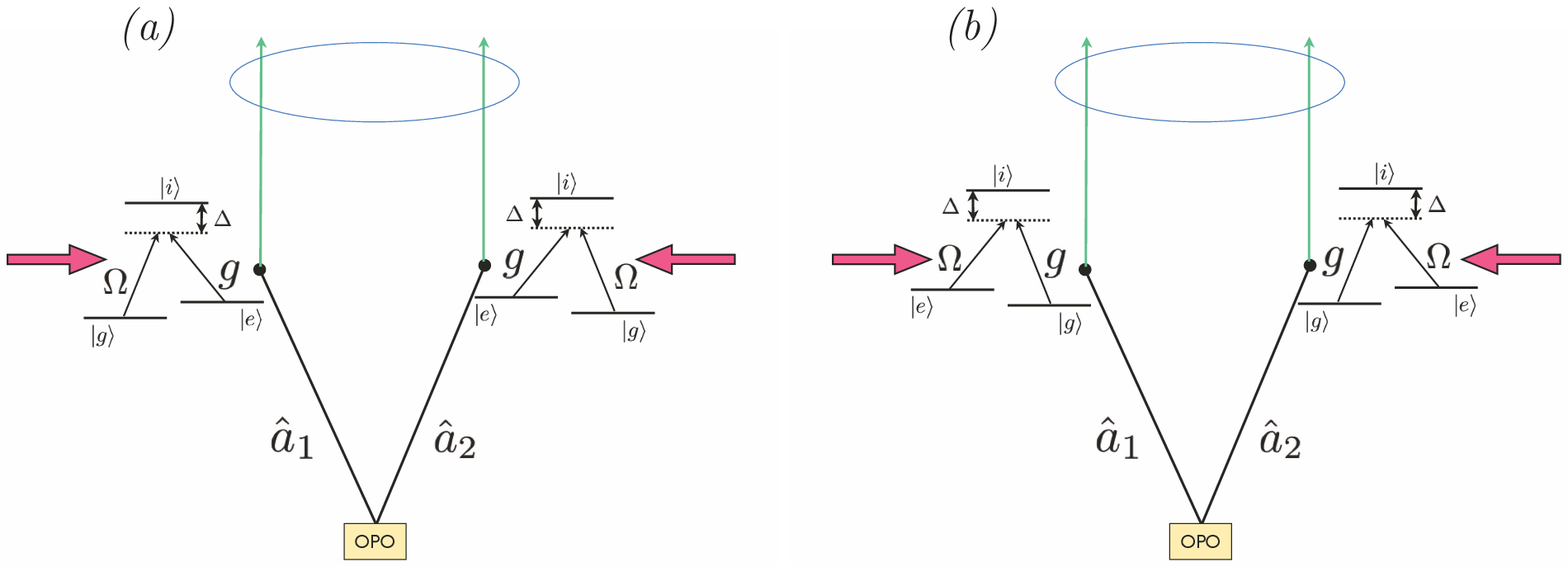}
\end{center}
\caption{Two-entangled light fields, generated for instance by an Optical Parametric Oscillator (OPO), impinge each on one atomic ensemble. The light field couples to one dipolar transition of a $\Lambda$-configuration of electronic levels (coupling strength $g$), while the second transition is driven by a laser (coupling strength $\Omega$). The light fields are here labeled by the corresponding photon-annihilation operators $a_1$ and $a_2$, while the involved atomic levels are $\ket{g}$ and $\ket{e}$ (stable states), with  $\ket{i}$ the common excited state. Before the interaction the atoms are prepared in state $\ket{g}$. The effective interaction between fields and collective atomic excitations is parametric for the coupling scheme shown in (a), namely, photonic and atomic excitations are simultaneously created or annihilated. In (b) the interaction is beam-splitter-like, i.e., the total number of photonic and atomic excitations is conserved. After the interaction, the collective state of the atomic ensembles is measured. As a result the entanglement between fields $a_1$ and $a_2$ is expected to increase (see text for details).} \label{fig1}
\end{figure}

The scheme for achieving CV entanglement purification of two optical modes is described in Fig.~\ref{fig1}. Two entangled light fields, with photon annihilation operators $a_1$ and $a_2$, propagates towards distant locations, where each field impinges onto an atomic ensemble. Each ensemble is homogeneously broadened and is constituted by a large number of identical atoms, whose relevant electronic transitions are composed by two stable states $\ket{g}$ and $\ket{e}$ coupled to a common excited state $\ket{i}$, and forming a $\Lambda$ configuration. The impinging field is far-off-resonantly coupled to the transition $\ket{g}\to\ket{i}$ ($\ket{e}\to\ket{i}$) with strength $g$, while a classical laser drives the transition $\ket{e}\to\ket{i}$ ($\ket{g}\to\ket{i}$) with coupling strength $\Omega$ (the two possible excitation models implement different dynamics, as discussed below). The parameters are chosen so that the excited state is effectively empty. The resulting effective interaction is a coherent two-photon coupling between the stable states $\ket{g}$ and $\ket{e}$, while spontaneous emission processes are negligible. In the following treatment we assume that density and size of the ensembles are essentially equal, so that the laser and input-field parameters can be taken to be the same for both atomic ensembles.

\subsection{Atom-Photon Interaction}

In this paper the interaction between atoms and light is assumed to be Hamiltonian. Denoting by $\rho$ the density matrix of optical fields and atoms, its evolution during the interaction is described by the von-Neumann equation
\begin{equation}
\label{vonNeumann}
\frac{\partial \rho}{\partial t}=\frac{1}{{\rm i}\hbar}[H,\rho]
\end{equation}
where the specific form of $H$ may take two forms (parametric or beam-splitter interaction) and is given below. For the schemes discussed in what follows, it is assumed that the atoms are all initially prepared in the internal state $\ket{g}$.

\subsubsection{The Parametric Interaction}

We first consider the configuration shown in Fig.~\ref{fig1}$(a)$. Here, the input field couples to the atomic transition $\ket{e} \rightarrow \ket{i}$, while the laser drives the transition  $\ket{g} \rightarrow \ket{i}$. In the regime in which the excited state is far-off resonance, it can be adiabatically eliminated and the dynamics of quantum field and atomic excitations can be described by the effective Hamiltonian~\cite{Duan01,Duan02}
\begin{eqnarray}
H = \sum_{l=1,2} \hbar\left(\frac{g\Omega_l}{\Delta} \sqrt{N_l}S_{eg}^{(l)} a_l^\dagger +{\rm H.c.}\right)
\label{eq:Hadiabatic}
\end{eqnarray}
where the index $l=1,2$ denotes the input optical mode and the atomic ensemble it impinges on, and $S_{eg}^{(l)}$ the collective spin operator for the corresponding ensemble. Here, $N_l$ is the atom number in ensemble $l$ and we denoted by $\Delta$ the detuning between the optical fields and the excited state. In Eq.(\ref{eq:Hadiabatic}), states $\ket{g}$ and $\ket{e}$ are resonantly coupled (ac-Stark shifts have been neglected as they can be compensated by suitably tuning the laser frequency).

In the Holstein-Primakoff representation~\cite{Holstein40}, the collective atomic operators (identified as angular momentum operators) can be associated with bosonic creation and annihilation operators via the relation $S^{(l)}_{ge} = (N_l-b_l^\dagger b_l)^{1/2} \, b_l/\sqrt{N_l}$, where $b_l$
denotes the bosonic annihilation operator satisfying the canonical commutation relations $[b_l,b_k^\dagger] = \delta_{lk}$. In the low excitation limit, where the number of atoms transferred to state $\ket{e}$ is small compared to the total number of atoms, the collective atomic operators can be approximated by $S^{(l)}_{ge} \approx b_l$, corresponding to taking the lowest order term in the expansion in powers of $\langle b_l^\dagger b_l\rangle/N_l$. By suitably choosing the phase of the coupling strengths $g$ and $\Omega_l$, the Hamiltonian~(\ref{eq:Hadiabatic}) can be reduced to the form
\begin{eqnarray}
H =  \sum_{l=1,2} i\hbar\mu_l (a_l^\dagger b_l^\dagger-a_l b_l),
\label{Eq:HeffParam}
\end{eqnarray}
where $\mu_l =  \sqrt{N_l}  |g\Omega_l|/\Delta$. Therefore, when the number of excited atoms is much smaller than the total number of atoms, the collective internal atomic excitation can be treated as continuous variable and the light-atom interaction for the setup of Fig.~\ref{fig1}$(a)$ is that of a parametric amplifier, in which a photon and a quantum of the atomic collective excitation are simultaneously created (annihilated). We shall assume from now on that $\mu_1=\mu_2=\mu$ (which could be realized by suitably adjusting the intensity of the two driving fields).


\subsubsection{The Beam Splitter Interaction}

We now consider the configuration shown in Fig.~\ref{fig1}$(b)$. The input optical field drives the $\ket{g} \rightarrow \ket{i}$ transition of the ensemble, while the laser couples to the transition  $\ket{e} \to \ket{i}$. Adiabatic elimination of the excited state gives a {\it beam splitter} form of the effective Hamiltonian, which reads
\begin{eqnarray}
H =  \sum_{l=1,2} i\hbar\nu_l (a_l^\dagger b_l-a_l b_l^\dagger),
\label{Eq:HeffBS}
\end{eqnarray}
where $\nu_l =   \sqrt{N_l}  |g\Omega_l|/\Delta$ and we made a suitable choice of the phase of the coupling strengths. Here the generation of a photon is accompanied by the \textit{absorption} of an atomic collective excitation. We assume also in this case that $\nu_1=\nu_2=\nu$.

\subsubsection{Discussion}

The two effective Hamiltonians introduced in this section determine the interaction between fields and atoms discussed in this paper. The
assumption of Hamiltonian dynamics is of course correct when incoherent processes can be neglected during the protocol. On the one hand, as we
already remarked, only stable atomic states are involved in the dynamics, while radiative decay can be neglected due to off-resonant excitation.
Moreover, it is also reasonable to neglect dephasing of the two metastable states which may arise from collisions between the atoms of the gas,
since for sufficiently dilute media the relevant coherence between the levels $\ket{g}$ and $\ket{e}$ can be maintained for a sufficiently long
time~\cite{Lukin00,Michael,Julsgaard01,Liu01,Lukin01}. One could avoid the effects of atomic collisions by choosing cold atoms, very dilute gases, or defects in solid matrices as rare-earth-doped solids, the latter suffering however from inhomogeneous broadening~\cite{Rare-Earth}. In such a case one has a smaller optical depth which, in our model, means smaller coupling constants $\mu$ and $\nu$, imposing longer interaction times. A further possibility is to employ nodes based on single atoms in traps, using the center-of-mass motion as bosonic mode. In this case large optical depth can be achieved by means of optical elements, such as a fiber or a cavity. In this context we note that the dynamics described by Hamiltonian~(\ref{Eq:HeffParam}) and~(\ref{Eq:HeffBS}) can be derived for this specific setup, where $a$ is here the cavity mode operator and $b$ the operator annihilating an excitation of the center-of-mass motion, see Refs.~\cite{Morigi,Morigi2,Canizares}.

For the case of atomic ensembles, an implicit but important physical assumption of our description is perfect spatial mode matching between each optical mode impinging on the atomic ensemble and the quantum optical modes naturally excited by the driven atomic medium in the $\Lambda$ configuration. Assuming perfect
mode matching means that the output field mode emerging from the atomic ensemble after the interaction (and the measurement) is exactly the same
optical mode entering the ensemble and therefore they are described by the same bosonic operators $a_j$. In general, however, the spatial
properties of the emerging field are determined by the geometric configuration of the gas cell and of the driving classical beams
\cite{Duan01,Duan02} and generally do not perfectly match with those of the input beams. Such limitation could be overcome with suitable optics
techniques. Significant results with entangled single photon beams have been recently achieved~\cite{Choi08}.

\subsection{The measurement}

After the interaction with the atomic ensemble, the second step leading to entanglement purification is a non-Gaussian operation. The easiest
way to implement it is to measure the state of the ancillary modes (the atomic collective modes), checking if a bosonic mode is in its vacuum
state or not. This consists in measuring whether each ensemble possesses or not spin excitations. In the case of an optical mode this projective
measurement is performed by single-photon on-off detectors which check if there is at least one photon in the mode. In an atomic system this
measurement can be implemented by suitably adapting the electron shelving or "quantum jump'' technique \cite{RMPWine} to this atomic
ensemble situation. Here, one applies a laser pulse, coupling state $\ket{e}$ with a fourth atomic level. If no fluorescence photon is observed it implies that state $\ket{e}$ is not populated, i.e., all the population resides in $\ket{g}$ (outcome $0$). If scattered photons are observed, some atomic excitations are present (outcome $1$). The detection procedure hence requires to distinguish no excitations from the presence of at least one excitation. For the atomic ensemble case, this requires measurement procedures like the one discussed in Refs.~\cite{imamoglu,kwiat}. When the node is a trapped ion, the procedure can be implemented by means of transferring the population of the ground state of the oscillator from the initial electronic state into a hide qubit and measuring the resonance fluorescence of the electronic state in a closed optical transition~\cite{Eschner}. The possible outcomes, $0$ ($1$) correspond to the projectors $\ket{0}_b\bra{0}$ ($\mathbb{I}_b - \ket{0}_b\bra{0}$). The projectors, corresponding to the measurement result, are described by the POVM elements
\begin{subequations}
\begin{eqnarray}
\{ E_{x_1=0},\ E_{x_1=1} \} &\equiv& \{ \ket{0}_{b_1}\bra{0},\ \mathbb{I}_{b_1}-\ket{0}_{b_1}\bra{0} \}\,,  \\
\{ E_{x_2=0},\ E_{x_2=1} \} &\equiv& \{ \ket{0}_{b_2}\bra{0},\ \mathbb{I}_{b_2}-\ket{0}_{b_2}\bra{0} \}\,,
\end{eqnarray}
\label{povm}
\end{subequations}
where $\mathbb{I}_{b_j}$ is the identity operator for the collective spin in ensemble $j$ and $\ket{0}_{b_1}\bra{0}$ the projector into the ground state (absence of excitations), which here corresponds to all atoms in state $|g\rangle$. The implementation of this step represents the non-Gaussian operation which may allow for purifying entanglement.  Denoting by $\bm{x} = (x_1,\ x_2)$ the generic outcome, the resulting density operator for modes $a_1$ and $a_2$ conditioned to the outcome $\bm{x}$ can be written as
\begin{equation}
\rho_{\bm{x}} = \frac{1}{p_{\bm{x}}} \textrm{Tr}_{b_1b_2}\{\rho E_{\bm{x}} \},
\label{rho:cond}
\end{equation}
where $p_{\bm{x}} = \textrm{Tr}_{a_1a_2} \{\rho_{\bm{x}}\}$ is the probability of the measurement outcome and $\textrm{Tr}_{b_1b_2}$ represents the trace over the atomic degrees of freedom. The (conditional) density matrix $\rho_{\bm{x}}$ describes now the state of the fields. The unconditional state, obtained without selecting the outcome, is $\rho' = \sum_{\bm{x}}  p_{\bm{x}} \rho_{\bm{x}}$. In this paper we will compare the entanglement of the fields described by the conditional state $\rho_{\bm{x}}$ with the initial state of the fields, checking whether the initial entanglement has been increased by the procedure.


\section{Time Evolution of the System and non-Gaussian Measurement}
\label{Sec:Measurement}

We now calculate the time evolution of the coupled atomic-optical modes and then determine the various conditional states after the measurement. We choose to work with the normally-ordered characteristic function which is defined as~\cite{WallsMilburn,Pirandola03}
\begin{equation}
\chi (\vec{\eta},t) = \textrm{Tr}\left\{ \rho(t) \prod_i e^{\eta_i\sigma_i^\dagger} \prod_i e^{-\eta_i^*\sigma_i} \right\},
\label{eq:charfun}
\end{equation}
with $\vec{\sigma} = (a_1,a_2,b_1,b_2)$ representing the set of bosonic operators, and $\vec{\eta} =
(\alpha_1,\alpha_2,\beta_1,\beta_2)$ with $\alpha_j,\beta_j$ complex numbers.

We shall consider a specific initial state of the optical fields, namely, a two-mode squeezed thermal state, while all the atoms are in the state $\ket{g}$, or equivalently in the ground state of the ensemble collective excitations. This yields the following form of the corresponding characteristic function
\begin{equation}
\label{Eq:chi0}
\chi(\vec\eta,0) = \exp{ \left\{ -\left(\frac{\lambda^2}{1-\lambda^2}+n_{th}\right)\left(|\alpha_1|^2+|\alpha_2|^2\right) \right\} } \exp{ \left\{ \frac{\lambda}{1-\lambda^2}\left(\alpha_1\alpha_2+\alpha_1^*\alpha_2^*\right) \right\} },
\end{equation}
with $n_{th}$ number of thermal photons and $\lambda$ the two-mode squeezing parameter. For $n_{th} = 0$ the state for the input fields becomes a standard two-mode squeezed state. The equation of motion for the characteristic function is obtained by taking the Fourier transform of the von Neumann equation~(\ref{vonNeumann}), see~\cite{WallsMilburn}, and for the parametric and beam-splitter case it takes the form
\begin{eqnarray}
\frac{\partial \chi_{par}(\vec{\eta},t)}{\partial t} &=& \mu \left( \alpha_1\beta_1 + \alpha_1^*\beta_1^* + \alpha_2\beta_2 +
\alpha_2^*\beta_2^*  -\alpha_1\frac{\partial}{\partial\beta_1^*} - \alpha_1^*\frac{\partial}{\partial\beta_1} -
\alpha_2\frac{\partial}{\partial\beta_2^*} - \alpha_2^*\frac{\partial}{\partial\beta_2} \right. \nonumber \\
&&\left.  - \beta_1\frac{\partial}{\partial\alpha_1^*} - \beta_1^*\frac{\partial}{\partial\alpha_1} -
\beta_2\frac{\partial}{\partial\alpha_2^*} - \beta_2^*\frac{\partial}{\partial\alpha_2} \right) \chi_{par}(\vec{\eta},t).
\label{eq:chi2redParam}\\
\frac{\partial \chi_{BS}(\vec{\eta},t)}{\partial t} &=& \nu \left( \alpha_1\frac{\partial}{\partial\beta_1} +\alpha_1^*\frac{\partial}{\partial\beta_1^*} + \alpha_2\frac{\partial}{\partial\beta_2} + \alpha_2^*\frac{\partial}{\partial\beta_2^*} \right. \nonumber \\
&&\left. - \beta_1\frac{\partial}{\partial\alpha_1} - \beta_1^*\frac{\partial}{\partial\alpha_1^*} - \beta_2\frac{\partial}{\partial\alpha_2} - \beta_2^*\frac{\partial}{\partial\alpha_2^*} \right) \chi_{BS}(\vec{\eta},t).
\label{eq:chi2redBS}
\end{eqnarray}
where we used the Hamiltonian in Eq.~(\ref{Eq:HeffParam}) and Eq.~(\ref{Eq:HeffBS}), respectively. The solutions of these equations are presented in Appendices~\ref{Sec:Chi-Param} and~\ref{Sec:ChiBS} and provide a full description of the state of the whole system of four modes, an optical and an atomic excitation mode at each separated node, after an interaction time $t$. Notice that the duration of the light-ensemble interaction is controlled by turning on and off the two driving laser fields.

The state of optical fields and atomic excitations, which evolves from the two-mode squeezed thermal state, Eq.~(\ref{Eq:chi0}) according to Eq.~(\ref{eq:chi2redParam}) or Eq.~(\ref{eq:chi2redBS}), is a Gaussian state of the four modes, since the initial state is Gaussian and the coherent evolution preserves the Gaussian nature. In order to increase the entanglement of the optical modes, hence, a non-Gaussian element must be introduced in the manipulation of the relevant optical modes~\cite{Eisert02,Browne03}. For the measurement described by the POVM in  Eq.~(\ref{povm}), where the outcome is either positive (presence of atomic excitations in the ensembles after the interaction) or negative (all atoms in the initial state), only the positive outcomes lead to non-Gaussian states of the optical modes. We will then focus on these results in order to check whether entanglement has increased with respect to the input state.

We proceed by determining the characteristic function of the optical modes, resulting from the measurement. In order to to so, we first need the density matrix of the full system in terms of the characteristic function,
\begin{eqnarray}
\rho &=&  \int \frac{d^2\alpha_1}{\pi} \frac{d^2\alpha_2}{\pi}\frac{d^2\beta_1}{\pi} \frac{d^2\beta_2}{\pi} \chi (\alpha_1, \alpha_2, \beta_1, \beta_2) \times \nonumber \\
&& \times D^\dagger(\alpha_1)D^\dagger(\alpha_2)D^\dagger(\beta_1)D^\dagger(\beta_2) e^{ (-|\alpha_1|^2 -|\alpha_2|^2 -|\beta_1|^2 -|\beta_2|^2)/2 },
\label{Eq:rhoITOchi}
\end{eqnarray}
where $D(\alpha_j)=\exp\left\{\alpha_j a_j^{\dagger}-\alpha_j^* a_j\right\}$ denotes the displacement operator \cite{WallsMilburn}. We use this density matrix in order to determine the four characteristic functions, corresponding to the four conditional states in Eq.~(\ref{rho:cond}). The latter can be written as
\begin{subequations}
\label{Eq:chivsI}
\begin{eqnarray}
\chi_{00}(\alpha_1,\alpha_2) &=& \frac{1}{p_{00}}I(\alpha_1,\alpha_2;1,1), \\
\chi_{01}(\alpha_1,\alpha_2) &=&  \frac{1}{p_{01}}\left[I(\alpha_1,\alpha_2;1,0) - I(\alpha_1,\alpha_2;1,1)\right], \\
\chi_{10}(\alpha_1,\alpha_2) &=&  \frac{1}{p_{10}}\left[I(\alpha_1,\alpha_2;0,1) - I(\alpha_1,\alpha_2;1,1)\right], \\
\chi_{11}(\alpha_1,\alpha_2) &=&  \frac{1}{p_{11}}\left[I(\alpha_1,\alpha_2;0,0) - I(\alpha_1,\alpha_2;1,0) \right. \nonumber \\
&& \left. - I(\alpha_1,\alpha_2;0,1) + I(\alpha_1,\alpha_2,1,1)\right]\,
\end{eqnarray}
\end{subequations}
where the probabilities $p_{x_1x_2}$ can be obtained from the normalization condition
\begin{equation}
\chi_{x_1x_2}(\alpha_1=0,\alpha_2=0)=1\,.
\label{norm}
\end{equation}
In Eqs.~(\ref{Eq:chivsI}) we have used
\begin{equation}
I(\alpha_1,\alpha_2; u,v) = \int d(u\beta_1) d(v\beta_2) \chi (\alpha_1, \alpha_2, \beta_1, \beta_2) e^{-|\beta_1|^2 -|\beta_2|^2 },
\label{eq:integral}
\end{equation}
where the measures are defined as follows
\begin{subequations}
\begin{eqnarray}
d(u\beta_1) &=& \left\{\begin{array}{lcr}
d^2\beta_1 & \;&u=1\\
\pi\delta^2(\beta_1)d^2\beta_1 &\;& u=0
\end{array}\right.
, \\
d(v\beta_2) &=& \left\{\begin{array}{lcr}
d^2\beta_2 & \;&v=1\\
\pi\delta^2(\beta_2)d^2\beta_2 &\;& v=0
\end{array}\right.,
\end{eqnarray}
\end{subequations}
The expressions for the integrals appearing in Eqs.~(\ref{Eq:chivsI}) and the accompanying probabilities are given in
Appendices~\ref{Sec:IntegralsParam} and~\ref{Sec:IntegralsBS}.


\subsection{Evaluating the Entanglement: Fidelity of Teleportation}

We remark that Eq.~(\ref{eq:integral}) is a Gaussian, and correspondingly the characteristic function for the outcome $(0,0)$, Eq.~(\ref{Eq:chivsI}a), is a Gaussian state. The characteristic functions for the other outcomes, in which at least one excitation is detected in one ensemble, are clearly non-Gaussian and could hence possibly be states where the initial entanglement has been purified. We note that these conditioned, non-Gaussian states are not pure. For this kind of states the available entanglement measures are rather cumbersome. In order to quantify the entanglement of these states we adopt the teleportation fidelity~\cite{Olivares03} as operational measure. The teleportation fidelity quantifies entanglement in terms of resource for teleportation. Here, the conditional output state is assumed to be initially shared by Alice and Bob for the standard CV-teleportation protocol~\cite{Kimble_Teleportation,Kimble_T_2,Pirandola06} and the average fidelity $F$ for teleporting a coherent state is calculated. In terms of characteristic functions, when these are symmetrically ordered, the teleportation fidelity takes the form~\cite{Pirandola06}
\begin{eqnarray}
F=\int  \frac{d^2\xi}{\pi} |\Phi^{in}(\xi)|^2 [\Phi^{ch}(\xi^*,\xi)]^*
\end{eqnarray}
where $\Phi^{ch}(\xi^*,\xi)$ is the characteristic function for the shared bipartite channel, the optical fields at the output in our case, and $\Phi^{in}(\xi)=\exp(-|\xi|^2/2)$ the characteristic function of the coherent state to be teleported~\footnote{Since in the expression of the fidelity the characteristic function of the coherent state appears with the modulus squared, a possible displacement of a coherent state does not matter.}. Using the relation between symmetrically- and normal-ordered characteristic functions~\cite{Cahill69b}, and
\begin{eqnarray}
\Phi^{ch}(\xi^*,\xi) = \chi_{x_1x_2}(\alpha_1=\xi^*,\alpha_2=\xi)\exp(-|\xi|^2)
\end{eqnarray}
for the shared bipartite channel, we arrive at the expression
\begin{eqnarray}
\label{Fidelity}
F_{x_1x_2}=\int  \frac{d^2\xi}{\pi} e^{-2|\xi|^2} \chi_{x_1x_2}(\alpha_1=\xi^*,\alpha_2=\xi).
\end{eqnarray}
Inserting the explicit expressions of the normally ordered characteristic function of the four conditional states in the right-hand-side of Eq.~(\ref{Fidelity}), after performing the integral we obtain the four conditional teleportation fidelities,
\begin{subequations}
\begin{eqnarray}
F_{00} &=& \frac{1}{p_{00}}\times \frac{1}{(B+1)^2-B_{12}^2}\times \frac{1}{2+2A-2A_{12}-2\frac{(C+D)^2}{B+1-B_{12}}}, \\
F_{01}&=& F_{10} = \frac{1}{p_{01}} \left[ \frac{1}{B+1}\times \frac{1}{2+2A-2A_{12}-\frac{(C+D)^2}{B+1}}\right. \nonumber \\
&&\left.\hspace{2.8cm}- \frac{1}{(B+1)^2-B_{12}^2}\times \frac{1}{2+2A-2A_{12}-2\frac{(C+D)^2}{B+1-B_{12}}}\right], \\
F_{11}&=& \frac{1}{p_{11}}\left[\frac{1}{2+2A-2A_{12}}-\frac{2}{B+1}\times \frac{1}{2+2A-2A_{12}-\frac{(C+D)^2}{B+1}}\right.\nonumber\\
&&\left.\hspace{2.8cm}+\frac{1}{(B+1)^2-B_{12}^2}\times \frac{1}{2+2A-2A_{12}-2\frac{(C+D)^2}{B+1-B_{12}}} \right],
\end{eqnarray}
\label{fidelities}
\end{subequations}
where the coefficients are given in Appendix A and B for the case of parametric and beam-splitter interaction, respectively (see also Appendices C and D for their final explicit expressions). CV-entanglement purification is achieved if one of these conditional fidelity is larger than the teleportation fidelity, evaluated when the bipartite channel is the initial state~(\ref{Eq:chi0}), and given by
\begin{equation}
F_{init} = \frac{1}{2}\frac{1-\lambda^2}{1-\lambda+n_{th}(1-\lambda^2)} .
\label{Eq:FidInit}
\end{equation}
We will denote it by {\it initial teleportation fidelity}.


\section{Results}

\begin{figure}[!t]
\begin{center}
\includegraphics[scale = 0.65]{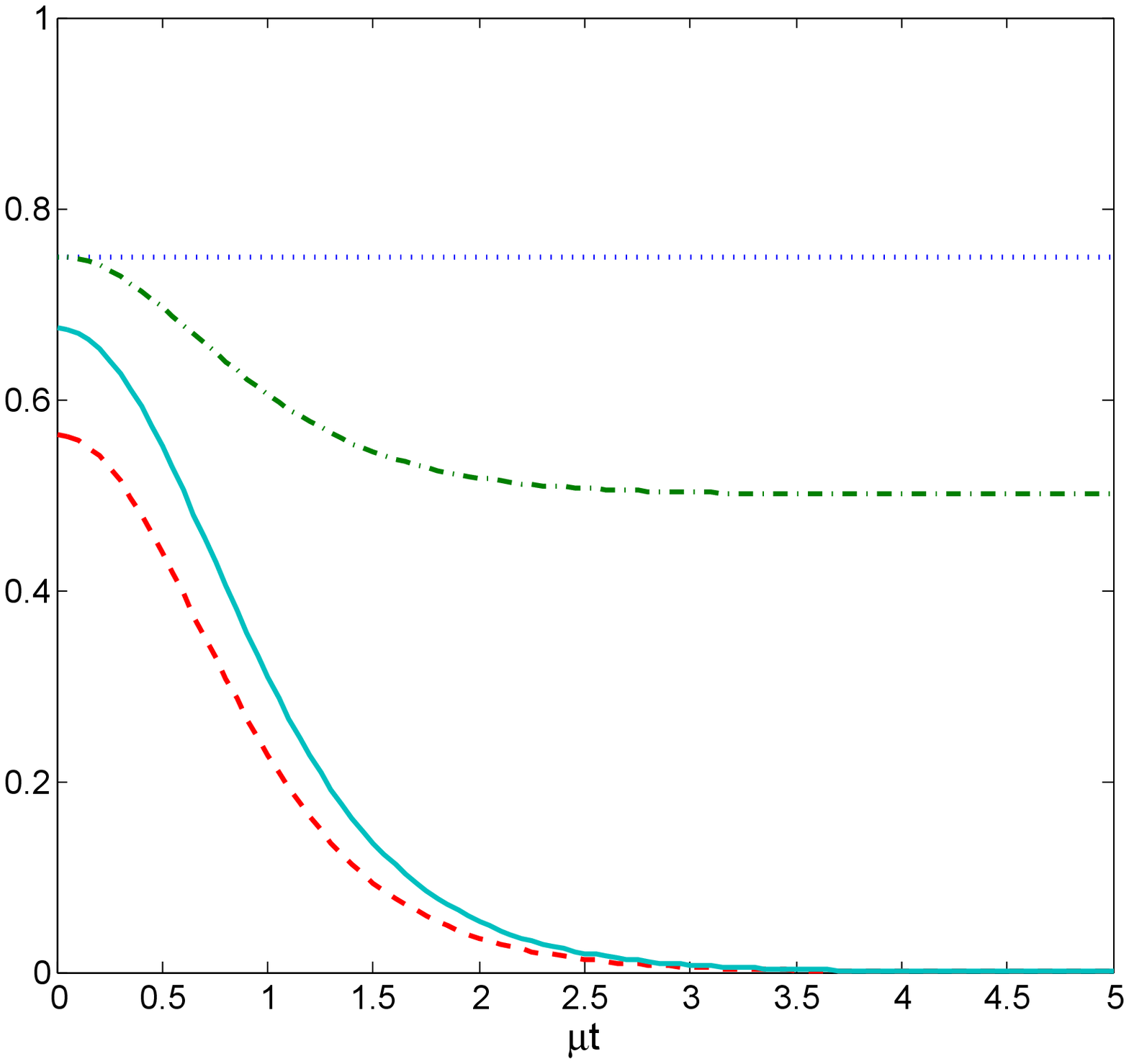}
\vspace{-2cm}
\caption{Fidelities for the four conditional states, obtained after parametric interaction with the atomic ensemble from Eqs.~(\ref{fidelities}), versus the interaction time $t$ (in units of $1/\mu$). Here, $\lambda = 0.5$ and $n_{th}=0$. The curves correspond to $F_{00}$ (dot-dashed line), $F_{01}=F_{10}$ (dashed line) and $F_{11}$ (solid line). The initial teleportation fidelity $F_{init}$ (dotted line) is shown for comparison. } \label{fig2}
\end{center}
\end{figure}

\begin{figure}[t]
\begin{center}
\includegraphics[scale = 0.75]{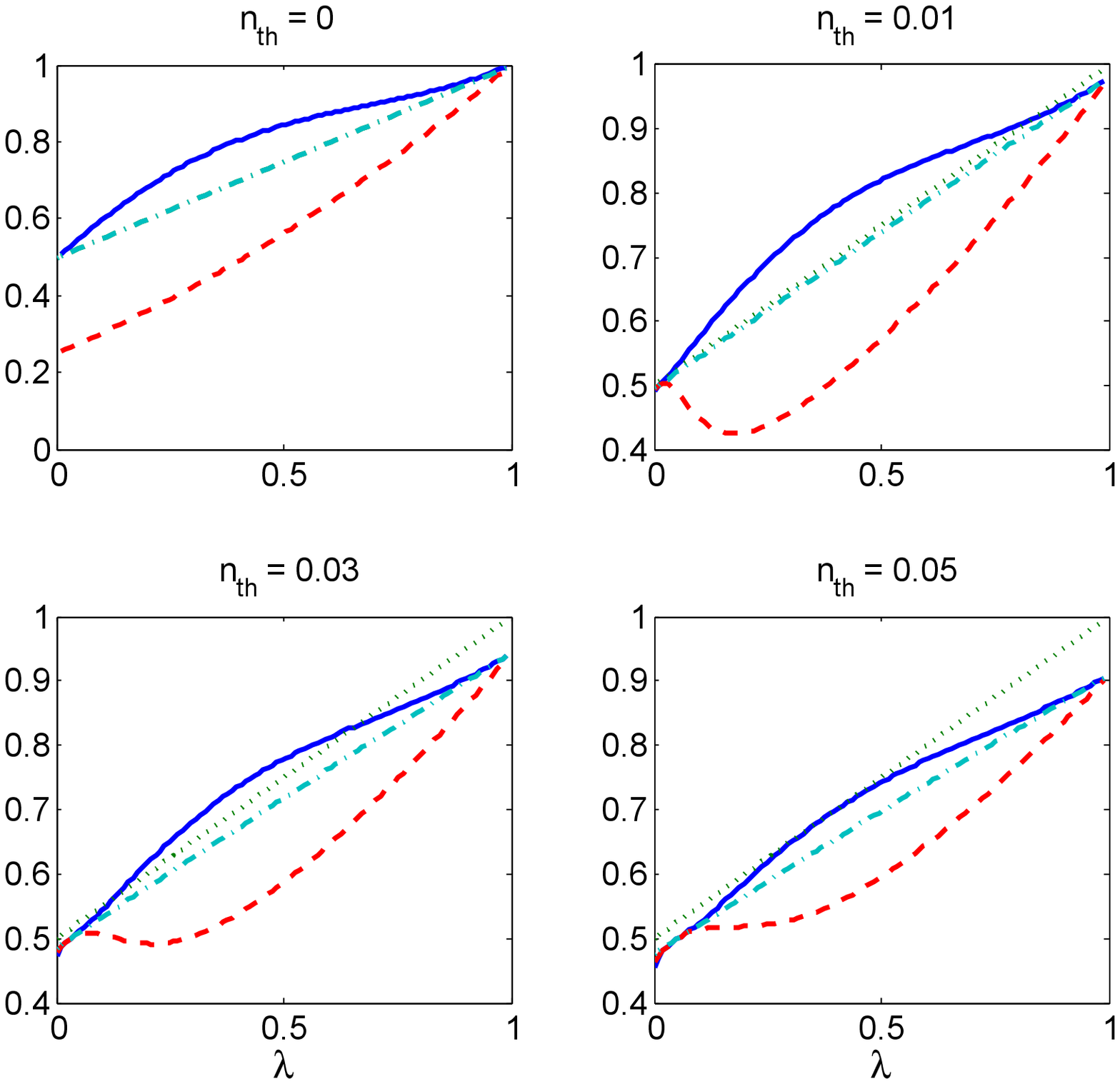}
\vspace{-5cm}
\caption{Fidelities for the beam-splitter interaction, as in Eqs.~(\ref{fidelities}), as a function of $\lambda$, and for various values of $n_{th}$. The fidelities are evaluated at the optimal time $t\sim \pi/\nu$, where $F_{11}$ is maximal. The curves correspond to $F_{00}$ (dot-dashed line), $F_{01}=F_{10}$ (dashed line) and $F_{11}$ (solid line). The initial teleportation fidelity $F_{init}$ (dotted line) is shown for comparison. }
\label{fig3}
\end{center}
\end{figure}

\begin{figure}[t]
\begin{center}
\includegraphics[scale = 0.75]{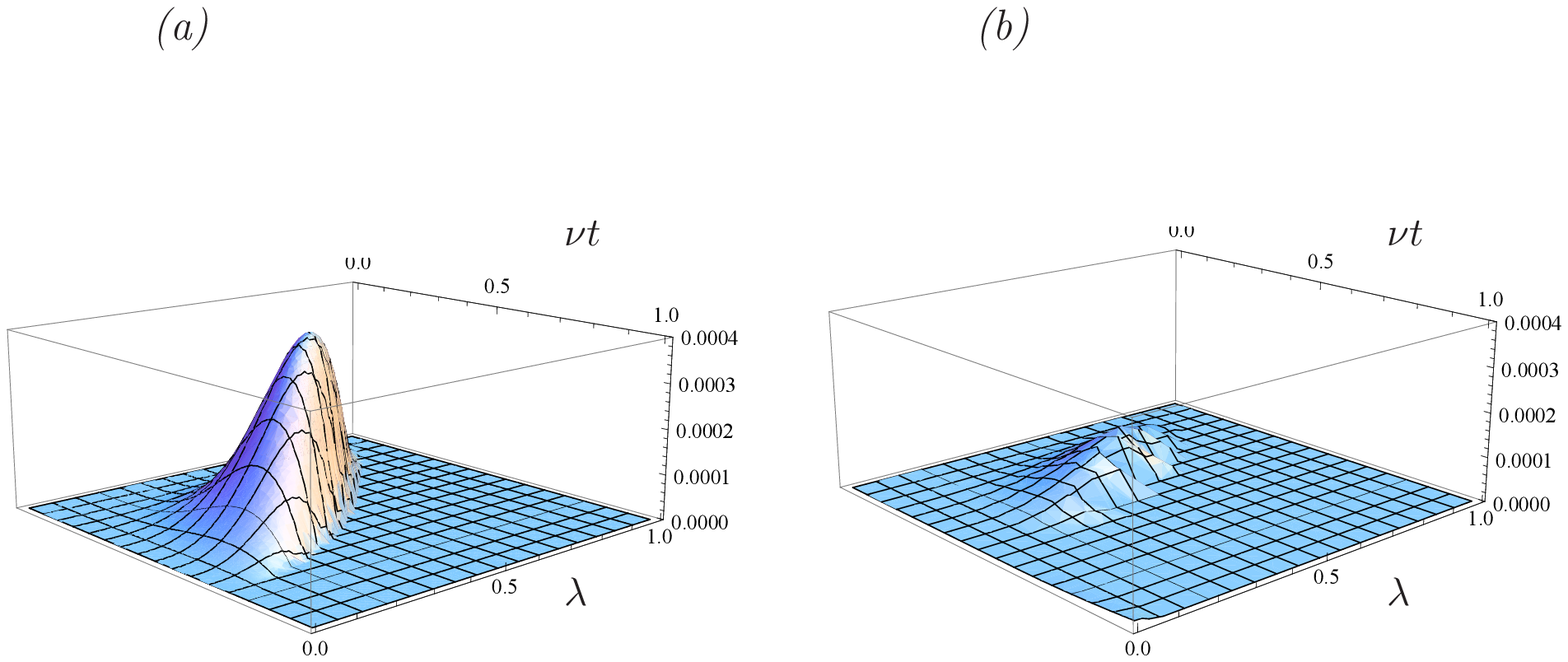}
\end{center}
\vspace{0.5cm}
\caption{Efficiency ${\mathcal E}$, Eq.~(\ref{eff}) as a function of $t$ (in units of $1/\nu$) and $\lambda$ for the beam-splitter interaction and $(a)$ $n_{th}=0$, $(b)$ $n_{th}=0.05$.
}
\label{fig4}
\end{figure}

We evaluate the teleportation fidelity for the parametric and the beam-splitter interaction between light and atoms, and study it as a function of the initial two-mode squeezing, characterized by the squeezing parameter $\lambda$, and by the number of thermal photon in the initial state $n_{th}$. We first consider the case of the parametric interaction. Figure~\ref{fig2} displays the four conditional fidelities versus the interaction time $t$, where the initial teleportation fidelity is shown for comparison. One clearly observes that none of the four conditional fidelities is larger than the initial one, implying that this protocol does not achieve entanglement purification when the interaction between light and atoms is parametric-like. This fact can be intuitively explained in terms of monogamy of entanglement \cite{Adesso04}: at each node the parametric interaction generates entanglement between the incident optical mode and the atomic collective mode, and this latter entanglement is detrimental for the all-optical entanglement we aim at increasing.
For the beam-splitter case, an analysis of the equations for the teleportation fidelities, Eqs.~(\ref{fidelities}), shows that $F_{00}$, $F_{01}$ and $F_{10}$ are always below $F_{init}$. On the contrary, $F_{11}$, the fidelity corresponding to the measurement outcome $11$, in which excitations are detected in both ensembles after the interaction can be greater than $F_{init}$ depending on both $t$ and $\lambda$ (besides $n_{th}$).
In particular there exist an optimal interaction time at which $F_{11}$ is maximal. It depends on $\lambda$ and $n_{th}$, and it is always close to $\pi/\nu$. Figure~\ref{fig3} displays the teleportation fidelities for the case of beam-splitter interaction as a function of $\lambda$ and $n_{th}$, for optimal time. One clearly observes that entanglement purification is achieved for $F_{11}$ when the initial value of $\lambda\approx 0.5$ and the number of thermal photons is small. As $n_{th}$ is increased, the range of values of $\lambda$ where entanglement purification is obtained decreases, till it vanishes. These results suggest to consider $F_{11}-F_{init}$ as figure of merit of the entanglement purification process.
A better estimate of the efficiency of the protocol should take into account the probability of the desired outcome, and can be made by means of the protocol \textit{efficiency} ${\mathcal E}$, which we define as
\begin{eqnarray}
{\mathcal E}=\left\{\begin{array}{ccc}
p_{11}\left(F_{11}-F_{init}\right)&\quad& \mathrm{if} \;\left(F_{11}-F_{init}\right)>0\\
0&\quad & \mathrm{if} \; \left(F_{11}-F_{init}\right)\le 0
\end{array}\right.
\label{eff}
\end{eqnarray}
The deviation of $F_{11}$ from $F_{init}$ is here weighted by the corresponding probability $p_{11}$ that the event occurs. Entanglement purification is achieved when $\mathcal E>0$.  Figure~\ref{fig4}(a) displays the protocol efficiency as a function of time and initial two-mode squeezing parameter $\lambda$ for $n_{th}=0$, showing that $\mathcal E$ is greater than zero in a wide region of parameters. In this case, for $n_{th}=0$ the initial state is pure, and in the corresponding region of parameters where $\mathcal E>0$ the protocol achieves entanglement \textit{concentration}. Figure~\ref{fig4}(b) displays the efficiency when $n_{th}>0$. Here, one sees that the protocol may achieve entanglement purification for appreciable intervals of interaction times $\nu t$ and initial two-mode squeezing $\lambda$. Clearly, the range of $n_{th}$ over which $\mathcal E$ remains greater than zero is finite.

If the roles of atomic and optical modes are exchanged in the measurement part of the protocol, i.e., one performs a measurement of presence or absence of photons in the \textit{optical} modes after the interaction, then the protocol achieves CV-entanglement between the atomic ensembles starting from entangled light modes and it hence performs a kind of entanglement swapping. Moreover, we find that in this way the entanglement of the ensembles conditional state, corresponding to the outcome $(1,1)$, is larger than the initial entanglement of the optical modes even in the case of parametric interaction, and in particular, it can be \textit{larger than} the entanglement that could be generated using an effective two-mode squeezing interactions between the ensembles.

\begin{figure}[!t]
\begin{center}
\includegraphics[scale = 0.75]{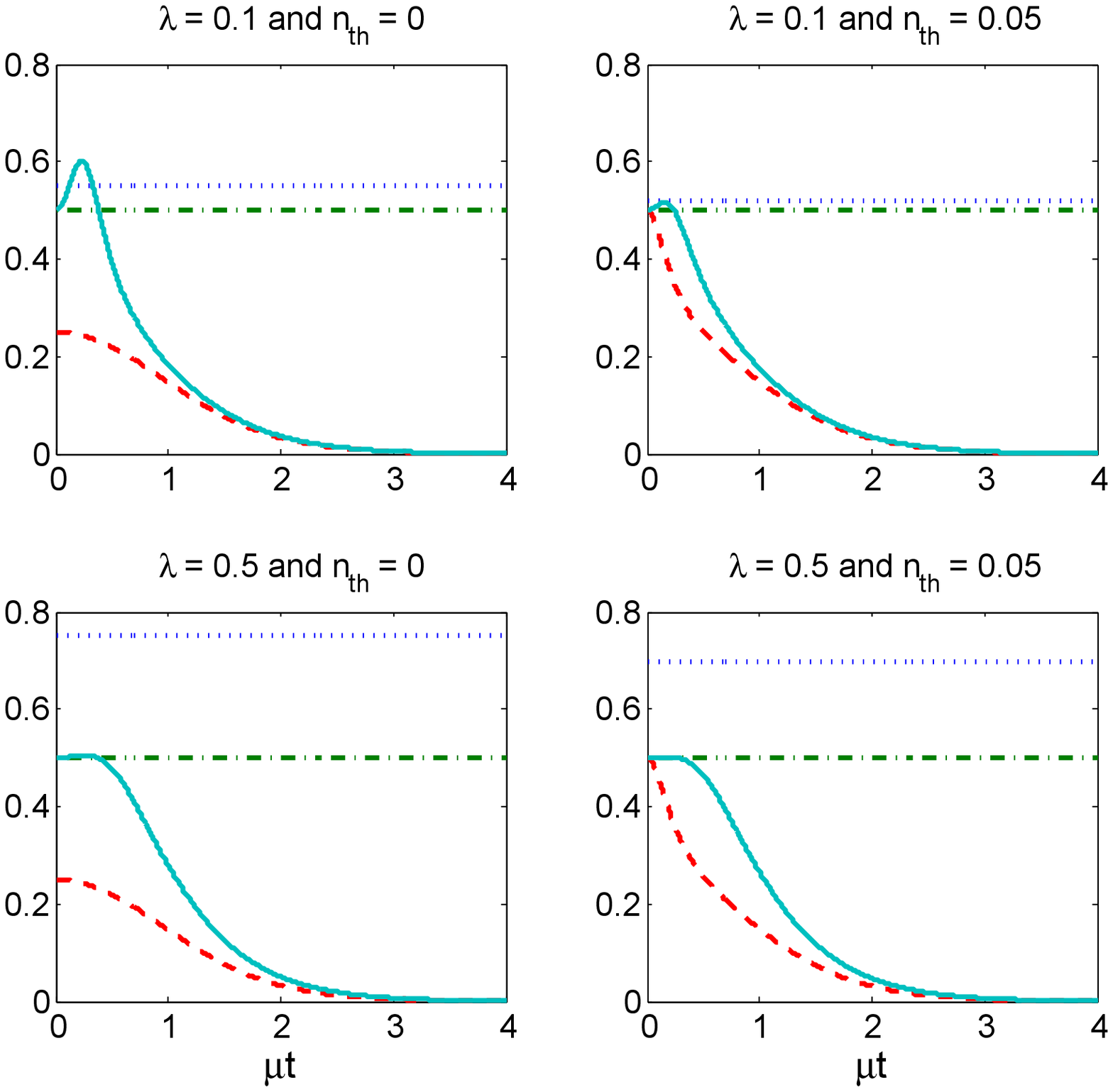}
\vspace{-5cm} \caption{Fidelities for entanglement swapping using the parametric interaction as a function of $\mu t$ for various values of $\lambda$ and $n_{th}$. The curves correspond to $F_{00}$ (dot-dashed line), $F_{01}=F_{10}$ (dashed line) and $F_{11}$ (solid line). The initial teleportation fidelity $F_{init}$ (dotted line) is shown for comparison.} \label{fig5}
\end{center}
\end{figure}

The calculations involved in this case mimic closely the calculations already performed in Sec~\ref{Sec:Measurement}, with the only difference of exchanging the roles of modes $a_i$ and $b_i$. This results in fidelities equivalent to those of Eqs.~(\ref{fidelities}), but with $A \leftrightarrow B$ and $A_{12} \leftrightarrow B_{12}$. Figure~\ref{fig5} displays the corresponding conditional teleportation fidelities for the parametric-amplifier interaction, showing an increase of $F_{11}$ with respect to $F_{init}$ for small values of $\lambda$ and short interaction times $t$, while it vanishes as time and $\lambda$ are increased.

\begin{figure}[!t]
\begin{center}
\includegraphics[scale = 0.8]{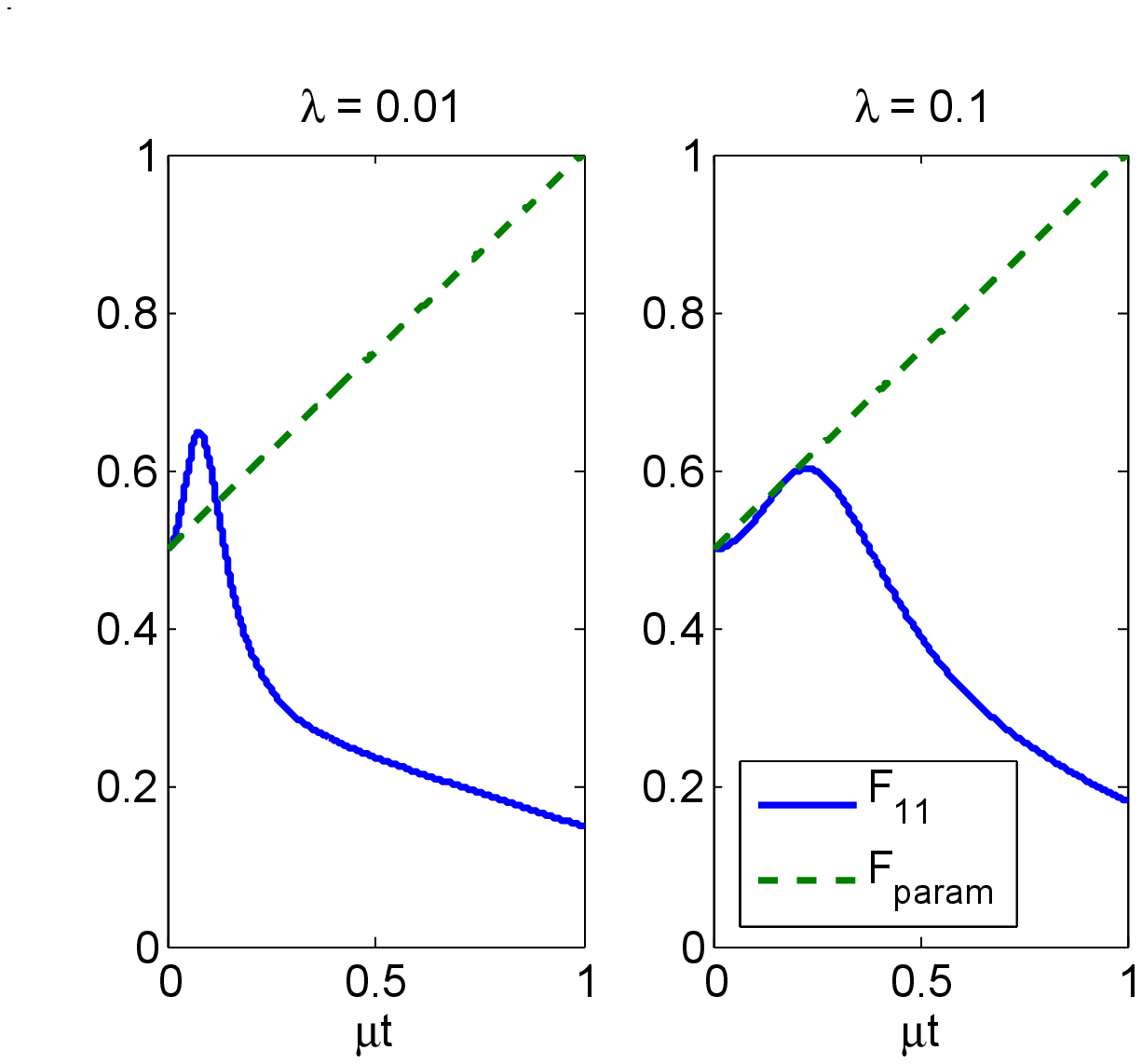}
\vspace{-5cm} \caption{Fidelity for entanglement swapping using the parametric interaction as a function of $\mu t$ for various values of $\lambda$. This fidelity is compared with the entanglement stemming from a pure parametric interaction, $F_{param}=(1+\mu t)/2$.} \label{fig6}
\end{center}
\end{figure}

Figure~\ref{fig6} displays $F_{11}$ versus time for small, fixed values of $\lambda$. The conditional fidelity is compared with the fidelity, when the entangled pair is generated by a parametric-amplifier Hamiltonian ($F_{param}=(1+\mu t)/2$). The increase is small, and vanishes as $n_{th}$ augments.

\section{Concluding Remarks}

A protocol for CV entanglement purification of two optical fields has been proposed. Entanglement purification is achieved by letting first the fields interact with an atomic node each, constituted for instance by the collective excitation of an atomic ensemble, and then by performing a dichotomic measurement of the presence of absence of atomic excitations on the atoms. We have considered two types of interactions between fields and atoms: parametric-like (where field and atomic excitations are simultaneously created or annihilated) and beam-splitter-like (where the total number of photons and atomic excitations is conserved). It turns out that by using the effective parametric amplifier interaction, it is impossible to improve the teleportation fidelity. This is an indication of entanglement monogamy in CV setting \cite{Adesso04}. On the contrary, by using the effective beam-splitter Hamiltonian we have shown that it is possible to enhance the input entanglement, even if the
initial weakly entangled state is not pure. From the perspective of this effective interaction, our protocol shares analogies with the protocol
for generation of entangled states through photon subtraction presented in Ref.~\cite{Kitagawa06,Takahashi09}.
In particular, we have extended the analysis to input thermal states and we have analyzed the performance in terms of a new global efficiency parameter which takes into account also the success probability of the protocol.
As a byproduct of our analysis we have shown that, if the role of the optical and atomic excitation mode are exchanged such that one performs the non-Gaussian measurement on the optical modes, one swaps entanglement between optical and atomic modes, thereby increasing the final entanglement of the atomic excitations over the initial one of the optical fields. In this case, we find a small improvement of the fidelity
even when the interaction is of the parametric-amplifier type.

The present protocol could be extended to the situation, in which the distant nodes are single atoms tightly confined by harmonic traps and inside high-Q resonators. In this case, the harmonic motion of the atomic center-of-mass plays the role of the ensemble excitation in the
protocol, and a beam-splitter or parametric interaction with the input, optical field can be tailored by using an external laser
field~\cite{Zeng,Parkins,Morigi,Morigi2,Canizares}. Measurement of whether the atom is in the trap ground state or not can be made by means of electron shelving techniques as used in laser cooling~\cite{Eschner}. An extensive discussion of the experimental requirements for a setup of this kind
can be found in~\cite{Morigi2,Canizares}.

In general, the use of atomic nodes in place of optical elements for entanglement purification of light fields provides one additional control tools, and our work is a first step in this direction.



\section{Acknowledgements}


This work was supported by the European Commission through the programmes FP6 IST-FET-QIPC projects QAP (Contract No. 015848) and SCALA (Contract No. 015714), and the FET-Open Project HIP (FP7-ICT-221899). SR acknowledges DEST ISL Grant No. CG090188 and ARC (DP0986932) for the financial support. GM acknowledges the support by the Spanish Ministerio de Ciencia y Innovaci\'on (Ramon-y-Cajal fellowship; Consolider Ingenio 2010 QOIT, CSD2006-00019; QNLP, FIS2007-66944) and by the German Research Council (Heisenberg professorship).

\begin{appendix}


\section{Characteristic Function - Parametric Interaction}
\label{Sec:Chi-Param}

Since the initial state is Gaussian and the involved Hamiltonian is bilinear, the generic state at time $t$ is again Gaussian and the general
solution of Eq.~(\ref{eq:chi2redParam}) can be written as
\begin{eqnarray}
\chi_{par}(\vec{\eta},t) &=& \textrm{exp} \left[ -A(|\alpha_1|^2 +|\alpha_2|^2) - B(|\beta_1|^2 + |\beta_2|^2)  \right.\nonumber \\
& &+ A_{12} \left. \left( \alpha_1\alpha_2+\alpha_1^*\alpha_2^* \right) + B_{12} \left( \beta_1\beta_2+\beta_1^*\beta_2^* \right) \right. \nonumber \\
&& \left. + C \left( \alpha_1\beta_2^*+\alpha_1^*\beta_2+\alpha_2\beta_1^*+\alpha_2^*\beta_1 \right)  \right. \nonumber \\
&& \left. + D \left( \alpha_1\beta_1+\alpha_1^*\beta_1^* + \alpha_2\beta_2+\alpha_2^*\beta_2^* \right)  \right] .
\label{eq:chi-sol-gen}
\end{eqnarray}
The individual coefficients in the exponent solve the set of six differential equations
\begin{subequations}
\begin{eqnarray}
&&\dot{A}=2\mu D,\\
&&\dot{B}=2\mu D,\\
&&\dot{D}=\mu (1+A+B),\\
&&\dot{A}_{12}=-2\mu C,\\
&&\dot{B}_{12}=-2\mu C,\\
&&\dot{C}=-\mu (B_{12}+A_{12}),
\end{eqnarray}
\end{subequations}
with the only nonvanishing initial conditions $A(0) = \frac{\lambda^2}{1-\lambda^2}+n_{th}$ and $A_{12}(0) = \frac{\lambda}{1-\lambda^2}$.
The solutions read
\begin{subequations}
\begin{eqnarray}
A(t) &=& \frac{\left[ 1+n_{th}(1-\lambda^2)\right] \cosh^2\mu t }{1 - \lambda^2}-1\\
B(t) &=& \frac{\left[ 1+n_{th}(1-\lambda^2)\right]\sinh^2\mu t}{1 - \lambda^2}\\
D(t) &=& \frac{1}{2}\frac{\left[ 1+n_{th}(1-\lambda^2)\right]\sinh (2\mu t)}{1 - \lambda^2}\\
C(t) &=& -\frac{\lambda}{2}\frac{\sinh (2\mu t)}{1 - \lambda^2} \\
A_{12}(t) &=& \lambda\frac{\cosh^2\mu t}{1 - \lambda^2}\\
B_{12}(t) &=& \lambda\frac{\sinh^2\mu t}{1 - \lambda^2}
\end{eqnarray}
\label{eq:chi-sol-coeffs-param}
\end{subequations}


\section{Characteristic Function - Beam Splitter Interaction}
\label{Sec:ChiBS}

Repeating the same argument of the preceding Section, the general solution of Eq.~(\ref{eq:chi2redBS}) for the beam splitter case can be written
as
\begin{eqnarray}
\chi_{BS}(\vec{\eta},t) &=& \textrm{exp} \left[ -A(|\alpha_1|^2 +|\alpha_2|^2) - B(|\beta_1|^2 + |\beta_2|^2)  + A_{12} \left( \alpha_1\alpha_2+\alpha_1^*\alpha_2^* \right) + B_{12} \left( \beta_1\beta_2+\beta_1^*\beta_2^* \right) \right. \nonumber \\
&& \left. + D \left( \alpha_1\beta_1^*+\alpha_1^*\beta_1+\alpha_2\beta_2^*+\alpha_2^*\beta_2 \right)
 + C \left( \alpha_1\beta_2+\alpha_1^*\beta_2^* + \alpha_2\beta_1+\alpha_2^*\beta_1^* \right)  \right].
\label{eq:chi-sol-genBS}
\end{eqnarray}
The individual coefficients solve the two sets of differential equations
\begin{subequations}
\begin{eqnarray}
\dot{A} &=& -2\nu D, \\
\dot{B} &=& 2\nu D, \\
\dot{D} &=& \nu (A-B),
\end{eqnarray}
and
\begin{eqnarray}
\dot{A}_{12} &=& 2\nu C, \\
\dot{B}_{12} &=& -2\nu C, \\
\dot{C} &=& -\nu (A_{12}-B_{12}),
\end{eqnarray}
\end{subequations}
with the only nonvanishing initial conditions $A(0) = \frac{\lambda^2}{1-\lambda^2}+n_{th}$ and $A_{12}(0) = \frac{\lambda}{1-\lambda^2}$. The
solutions now read
\begin{subequations}
\begin{eqnarray}
A(t) &=& \left[ \lambda^2+n_{th}(1-\lambda^2)\right]\frac{\cos^2 \nu t}{1-\lambda^2}, \\
B(t) &=& \left[ \lambda^2+n_{th}(1-\lambda^2)\right]\frac{\sin^2 \nu t}{1-\lambda^2}, \\
D(t) &=& \frac{1}{2}\left[ \lambda^2+n_{th}(1-\lambda^2)\right]\frac{\sin(2\nu t)}{1-\lambda^2}, \\
A_{12}(t) &=& \lambda  \frac{\cos^2 \nu t}{1-\lambda^2}, \\
B_{12}(t) &=& \lambda \frac{\sin^2 \nu t}{1-\lambda^2}, \\
C(t) &=& \frac{-\lambda}{2}\frac{\sin(2\nu t)}{1-\lambda^2}.
\end{eqnarray}
\label{eq:chi-sol-coeffs-BS}
\end{subequations}


\section{Fidelities - Parametric Interaction}
\label{Sec:IntegralsParam}

Using the results of Appendix~\ref{Sec:Chi-Param}, the integrals \eqref{eq:integral} can be easily evaluated giving the following results
\begin{subequations}
\begin{eqnarray}
I(\alpha_1,\alpha_2;0,0)&=&\exp\left[-A\left(|\alpha_1|^2+|\alpha_2|^2\right) +A_{12}\left(\alpha_1\alpha_2+\alpha_1^*\alpha_2^*\right)\right], \\
I(\alpha_1,\alpha_2;0,1)&=&\frac{1}{B+1}\exp\left[-\left(A-\frac{D^2}{B+1}\right)|\alpha_1|^2-\left(A-\frac{C^2}{B+1}\right)
|\alpha_2|^2\right. \nonumber \\
&& \left. +\left(A_{12}+\frac{CD}{B+1}\right)\left(\alpha_1\alpha_2+\alpha_1^*\alpha_2^*\right)\right], \\
I(\alpha_1,\alpha_2;1,0)&=&\frac{1}{B+1}\exp\left[-\left(A-\frac{C^2}{B+1}\right)|\alpha_1|^2-\left(A-\frac{D^2}{B+1}\right)
|\alpha_2|^2\right. \nonumber \\
&& \left. +\left(A_{12}+\frac{CD}{B+1}\right)\left(\alpha_1\alpha_2+\alpha_1^*\alpha_2^*\right)\right], \\
I(\alpha_1,\alpha_2;1,1)&=&\frac{1}{(B+1)^2-B_{12}^2}\exp\left[-\left(A-\frac{(B+1)(C^2+D^2)+2B_{12}CD}{(B+1)^2-B_{12}^2}\right)\left(|\alpha_1|^2+|\alpha_2|^2\right)\right.\nonumber\\
&&\left.\hspace{3.2cm}+\left(A_{12}+\frac{2(B+1)CD+B_{12}(C^2+D^2)}{(B+1)^2-B_{12}^2}\right)
\left( \alpha_1\alpha_2+\alpha_1^*\alpha_2^*\right)\right]. \nonumber\\
\end{eqnarray}
\end{subequations}
The probabilities are then derived using the condition (\ref{norm})
\begin{subequations}
\begin{eqnarray}
p_{00} &=& I(0,0;1,1)=\frac{1}{(B+1)^2-B_{12}^2}, \\
p_{01}&=&  I(0,0;1,0) - I(0,0;1,1)=\frac{1}{B+1}-\frac{1}{(B+1)^2-B_{12}^2}, \\
p_{10}&=&  I(0,0;0,1) - I(0,0;1,1)=\frac{1}{B+1}-\frac{1}{(B+1)^2-B_{12}^2}, \\
p_{11}&=&  I(0,0;0,0) - I(0,0;1,0) - I(0,0;0,1) + I(0,0;1,1) \nonumber \\
&=&1-\frac{2}{B+1}+\frac{1}{(B+1)^2-B_{12}^2}.
\end{eqnarray}
\label{chivsI}
\end{subequations}
Finally, with the help of Eqs.(\ref{Eq:chivsI}) and \eqref{Fidelity} we can easily arrive at Eq.\eqref{fidelities}.


\section{Fidelities - Beam Splitter Interaction}
\label{Sec:IntegralsBS}
Using the results of Appendix~\ref{Sec:ChiBS}, the integrals \eqref{eq:integral} can be easily evaluated giving the following result
\begin{subequations}
\begin{eqnarray}
I(\alpha_1,\alpha_2;0,0)&=&\exp\left[-A\left(|\alpha_1|^2+|\alpha_2|^2\right)+A_{12}\left(\alpha_1\alpha_2+\alpha_1^*\alpha_2^*\right)\right], \\
I(\alpha_1,\alpha_2;0,1)&=&\frac{1}{B+1}\exp\left[-\left(A-\frac{C^2}{B+1}\right)|\alpha_1|^2-\left(A-\frac{D^2}{B+1}\right)
|\alpha_2|^2 \right. \nonumber\\
&&\left. +\left(A_{12}+\frac{CD}{B+1}\right)\left(\alpha_1\alpha_2+\alpha_1^*\alpha_2^*\right)\right], \\
I(\alpha_1,\alpha_2;1,0)&=&\frac{1}{B+1}\exp\left[-\left(A-\frac{D^2}{B+1}\right)|\alpha_1|^2-\left(A-\frac{C^2}{B+1}\right)
|\alpha_2|^2\right. \nonumber\\
&&\left. +\left(A_{12}+\frac{C D}{B+1}\right)\left(\alpha_1\alpha_2+\alpha_1^*\alpha_2^*\right)\right], \nonumber\\
I(\alpha_1,\alpha_2;1,1)&=&\frac{1}{(B+1)^2-B_{12}^2}\exp\left[-\left(A-\frac{(B+1)(D^2+C^2)+2B_{12}C D}{(B+1)^2-B_{12}^2}\right)\left(|\alpha_1|^2+|\alpha_2|^2\right)\right.\nonumber\\
&&\left.+\left(A_{12}+\frac{2(B+1)C D+B_{12}(C^2+D^2)}{(B+1)^2-B_{12}^2}\right)
\left( \alpha_1\alpha_2+\alpha_1^*\alpha_2^*\right)\right]. \nonumber\\
\end{eqnarray}
\end{subequations}
The probabilities are then derived using the condition (\ref{norm})
\begin{subequations}
\begin{eqnarray}
p_{00} &=& I(0,0;1,1)=\frac{1}{(B+1)^2-B_{12}^2}, \\
p_{01}&=&  I(0,0;1,0) - I(0,0;1,1)=\frac{1}{B+1}-\frac{1}{(B+1)^2-B_{12}^2}, \\
p_{10}&=&  I(0,0;0,1) - I(0,0;1,1)=\frac{1}{B+1}-\frac{1}{(B+1)^2-B_{12}^2}, \\
p_{11}&=&  I(0,0;0,0) - I(0,0;1,0) - I(0,0;0,1) + I(0,0;1,1) \nonumber \\
&=&1-\frac{2}{B+1}+\frac{1}{(B+1)^2-B_{12}^2}.
\end{eqnarray}
\label{chivsIBS}
\end{subequations}
Finally, with the help of Eqs.(\ref{Eq:chivsI}) and \eqref{Fidelity} we can easily arrive at Eq.\eqref{fidelities}.

\end{appendix}

\newpage

\section*{Figure Captions}

{\bf Figure 1:} Two-entangled light fields, generated for instance by an Optical Parametric Oscillator (OPO), impinge each on one atomic ensemble. The light field couples to one dipolar transition of a $\Lambda$-configuration of electronic levels (coupling strength $g$), while the second transition is driven by a laser (coupling strength $\Omega$). The light fields are here labeled by the corresponding photon-annihilation operators $a_1$ and $a_2$, while the involved atomic levels are $\ket{g}$ and $\ket{e}$ (stable states), with  $\ket{i}$ the common excited state. Before the interaction the atoms are prepared in state $\ket{g}$. The effective interaction between fields and collective atomic excitations is parametric for the coupling scheme shown in (a), namely, photonic and atomic excitations are simultaneously created or annihilated. In (b) the interaction is beam-splitter-like, i.e., the total number of photonic and atomic excitations is conserved. After the interaction, the collective state of the atomic ensembles is measured. As a result the entanglement between fields $a_1$ and $a_2$ is expected to increase (see text for details).

{\bf Figure 2:} Fidelities for the four conditional states, obtained after parametric interaction with the atomic ensemble from Eqs.~(\ref{fidelities}), versus the interaction time $t$ (in units of $1/\mu$). Here, $\lambda = 0.5$ and $n_{th}=0$. The curves correspond to $F_{00}$ (dot-dashed line), $F_{01}=F_{10}$ (dashed line) and $F_{11}$ (solid line). The initial teleportation fidelity $F_{init}$ (dotted line) is shown for comparison.

{\bf Figure 3:} Fidelities for the beam-splitter interaction, as in Eqs.~(\ref{fidelities}), as a function of $\lambda$, and for various values of $n_{th}$. The fidelities are evaluated at the optimal time $t\sim \pi/\nu$, where $F_{11}$ is maximal. The curves correspond to $F_{00}$ (dot-dashed line), $F_{01}=F_{10}$ (dashed line) and $F_{11}$ (solid line). The initial teleportation fidelity $F_{init}$ (dotted line) is shown for comparison.

{\bf Figure 4:} Efficiency ${\mathcal E}$, Eq.~(\ref{eff}) as a function of $t$ (in units of $1/\nu$) and $\lambda$ for the beam-splitter interaction and $(a)$ $n_{th}=0$, $(b)$ $n_{th}=0.05$.

{\bf Figure 5:} Fidelities for entanglement swapping using the parametric interaction as a function of $\mu t$ for various values of $\lambda$ and $n_{th}$. The curves correspond to $F_{00}$ (dot-dashed line), $F_{01}=F_{10}$ (dashed line) and $F_{11}$ (solid line). The initial teleportation fidelity $F_{init}$ (dotted line) is shown for comparison.

{\bf Figure 6:} Fidelity for entanglement swapping using the parametric interaction as a function of $\mu t$ for various values of $\lambda$. This fidelity is compared with the entanglement stemming from a pure parametric interaction, $F_{param}=(1+\mu t)/2$.

\end{document}